\documentstyle[aps]{revtex}

\def\lta{~\raise.4ex\hbox{$<$}\llap{\lower.6ex\hbox{$\sim$}}~}
\def\gta{~\raise.4ex\hbox{$>$}\llap{\lower.6ex\hbox{$\sim$}}~}

\begin{document} \input psfig.sty  

\title{Cooperation and Self-Regulation in a Model of Agents Playing Different Games}
\author{H. Fort}

\address{Instituto de F\'{\i}sica, Facultad de Ciencias, Universidad de la
Rep\'ublica, Igu\'a 4225, 11400 Montevideo, Uruguay}

\maketitle

\begin{abstract}

A simple model for cooperation between "selfish" agents, which play an extended 
version of the Prisoner's Dilemma(PD) game, in which they use arbitrary payoffs, 
is presented and studied. A continuous variable, representing the probability of 
cooperation, $p_k(t) \in$ [0,1], is assigned to each agent $k$ at time $t$. At 
each time step $t$ a pair of agents, chosen at random, interact by playing the 
game. The players update their $p_k(t)$ using a criteria based on the 
comparison of their utilities with the simplest estimate for expected income. 
The agents have no memory and use strategies not based on direct reciprocity nor 
'tags'. 
Depending on the payoff matrix, the systems self-organizes - after a transient -  
into stationary states characterized by their average probability of cooperation 
$\bar{p}_{eq}$ and average equilibrium per-capita-income 
$\bar{p}_{eq},\bar{U}_\infty$.
It turns out that the model exhibit some results that contradict the intuition. 
In particular, some games which - {\it a priory}- seems to favor defection most, 
may produce a relatively high degree of cooperation. Conversely, other games, 
which one would bet that lead to maximum cooperation, indeed are not the
optimal for producing cooperation.     
\end{abstract}

\vspace{2mm}

{\it keybords}: Complex adaptive systems, Agent-based models, Social systems

\vspace{1mm}


\vspace{2mm}

\section{Introduction}

Game Theory constitutes a powerful and versatile approach to analyze the collective 
behavior of adaptive agents, from humans to bacteria and firms.   
In particular, the {\it Prisoner's Dilemma} (PD) game plays in Game Theory a role 
similar to the harmonic oscillator in Physics. It's been also referred to as the 
{\it E. Coli} of Social Sciences, allowing a very large variety of studies. Indeed,
this game, developed in the early fifties, offers a very simple and intuitive 
approach to the problem of how cooperation emerges in societies of "selfish" 
individuals {\it i.e.} individuals which pursue exclusively their own self-
benefit. It was used in a series of works by Robert Axelrod and co-workers 
\cite{axel84} to examine the basis of cooperation between selfish agents in a 
wide variety of contexts. 
Furthermore, mechanisms of cooperation based on the PD have shown their 
usefulness in Political Science \cite{h90}-\cite{g88}, Economics \cite{ff02}-
\cite{w89}, International Affairs  \cite{h01}-\cite{s71}, 
Theoretical Biology \cite{wn99}-\cite{n90} and Ecology \cite{md97}-
\cite{dmh92}. 

The PD game consists in two players, say $i$ and $j$, each confronting two 
choices: cooperate (C) or defect (D) and each makes its choice without
knowing what the other will do. 
The four possible outcomes for the interaction of agent $i$ with agent $j$ are
: 1) they can both cooperate (C,C) 2) both defect (D,D), 3) $i$ cooperates and 
$j$ defects(C,D) and 4) $i$ defects and $j$ cooperates (D,C). 
Depending on the situation 1)-4), the agent $i$ ($j$) gets respectively
: the "reward" $R (R)$, the "punishment" $P (P)$, the "sucker's 
payoff" $S$ (the "temptation to defect" $T$) or $T (S)$.
These four payoffs obey the following chain of inequalities: 
\begin{equation}
T>R>P>S;
\label{eq:chainPD}
\end{equation}
for instance the four canonical PD payoffs are: $R=3, S=0, T=5$ and $P=1$.
Clearly it pays more to defect: if one of the two players defects -say $i$- , 
the other who cooperates will end up with nothing. In fact, even if agent $i$ 
cooperates, agent $j$ should defect, because in that case he will get $T$ which 
is larger than $R$. That is,  
independently of what the other player does, defection D yields a higher payoff 
than cooperation and is the {\it dominant strategy} for rational agents.
This is equivalent to say, in a more technical language that, the outcome (D,D) 
of both players is the Nash equilibrium \cite{n51} of the PD game.
The dilemma is that if both defect, both do worse than if
both had cooperated: both players get $P$ which is smaller than $R$.

One can assign a {\em payoff matrix} M$^{RSTP}$ to the PD game given by 

\vspace{-4mm}

\begin{center}

$${\mbox M}^{RSTP}=\left(\matrix{(R,R)&(S,T)\cr (T,S)&(P,P) \cr}\right),$$

\end{center}
which summarizes the payoffs for {\it row} actions when confronting
with {\it column} actions.

The emergence of cooperation in prisoner's dilemma (PD) games is generally 
assumed to require repeated play (and strategies such as 
{\it Tit for Tat} (TFT) \cite{axel84}, 
involving memory of previous interactions) or features ("tags") 
permitting cooperators and defectors to distinguish one another 
\cite{ep98}.

In this work, I consider a simple model of selfish agents, possessing 
neither memory nor tags, to study the self-organized cooperative states 
which emerge when they play an {\it extended} PD game with arbitrary payoffs, 
{\it i.e.} payoffs which do not necessarily fulfill inequalities 
(\ref{eq:chainPD}). The taxonomy of
2x2 games (one-shot games involving two players with two actions each) was
constructed by Rapoport and Guyer \cite{rg66}, 
and showed that
there exist exactly 78 non-equivalent games.

There are  $N_{ag}$ agents, with one variable assigned to each agent at the site 
or cell $k$ and at time $t$: his probability of cooperation $p_k(t)$. Pairs of 
agents, $i$ and $j$, interact by playing the PD game at 
each time step $t$.
I use a Mean Field (MF) approach, in which all the spatial correlations 
in the system are neglected, and thus agent $i$ and $j$ are chosen at
random. After playing the PD the players update their probability of cooperation 
$p_i(t)$ and $p_j(t)$ according to the same definite "measure of success" 
which does not vary with time. Thus all agents follow a universal and 
invariant strategy defined by a measure of success plus an updating 
rule to transform $p_i(t)$ and $p_j(t)$ into $p_i(t+1)$ and $p_j(t+1)$. 

After a transient, the system self-organizes into a state of equilibrium
characterized by the average probability of cooperation $\bar{p}_{eq}$
which depend on the payoff matrix.

Payoff matrices can be classified into sub-categories 
according to their dominant strategy. 
Let us call $M_D$ the class of those matrices such that:
\begin{equation}
T>R, \;\; \mbox{and} \;\; P>S,
\label{eq:class1}
\end{equation}
for which the dominant strategy is D. This class comprises, for instance, the 
canonical matrix M$^{3051}$ and M$^{1053}$, etc.
A second class $M_C$ corresponds to 
\begin{equation}
R>T, \;\; \mbox{and} \;\; S>P,
\label{eq:class2}
\end{equation}
for which the dominant strategy is C, examples of this class are 
the matrices: M$^{5310}$
and M$^{3501}$. 
The remaining matrices do not comply with
equation (\ref{eq:class1}) or (\ref{eq:class2}) and produce situations
, {\it a priori}, not dominated by (D,D) or (C,C).

One might wonder why bother to study matrices which imply no dilemma and are 
unrealistic 
in order to model the social behavior of the majority
of individuals. There are several reasons.
First, this "unreasonable" payoff matrices 
can be used by minorities of individuals which depart from the "normal" ones 
(assumed to be neutral). For instance, "antisocial" "always D" individuals, 
which cannot 
appreciate any advantage of cooperation, or 
"altruistic" "always C" individuals.
Second, it seems interesting to test the robustness of cooperation under 
changes in the payoff matrix. In particular, we will se that even payoff 
matrices which 
imply a dilemma can produce either $\bar{p}_{eq}=0.5$ or 
$\bar{p}_{eq}=0$. Third, arbitrary payoff matrices could be also of 
importance in other contexts different from societies. One might envisage 
situations 
in which a definite value of $\bar{p}_{eq}$ is required or is desirable in the 
design
of a system or is the one which optimizes the functioning of a 
particular mechanism, etc. For example, to understand how a market of competing 
firms attains self-regulation. Or for instance in the traffic problem, where the 
damage suffered 
from mutual D (crash) exceeds the damage suffered by being exploited (turn 
away), which is more appropriately described by the so-called {\it chicken game} 
for which $T>R>S>P$.

Fourth, we will show results for these payoff matrices which, at first
glance, defy our intuition. For example, payoff matrices which, at least in
principle, one would bet that favor defection and indeed produce a  
not so low degree of cooperation.

\section{A Mechanism to produce Cooperative Equilibrium States }

Among the weaknesses of major approaches that have been 
considered to answer the question about the emergence of cooperation
two are often remarked. 
The first criticism is about the over-simplification in the behavior of agents: 
they either always cooperate (C) or always defect (D). Clearly, this is not very 
realistic. Indeed, the levels of cooperation of the individuals seem to exhibit
a continuous gamma of values. 
The second objection is concerning the deterministic nature of the algorithms 
which seem to fail to incorporate the stochastic component of human behavior.

Both problems can be overcome by assigning to each agent $k$ a probability of 
cooperation $p_k(t)$ (a real number in the interval [0,1]) instead of definite 
behaviors like C or D. Concerning the first objection, $p_k(t)$  reflects the 
existence of a "gray scale" of levels of cooperation instead of just "black" and 
"white". Regarding the second objection, 
the proposed algorithm is clearly non deterministic: agent $k$ plays C with 
probability $p_k$ and D with probability 1-$p_k$.

Now, let us describe the dynamics.
The pairs of interacting partners, by virtue of the MF treatment,
are chosen randomly instead of being restricted to some 
neighborhood. The implicit assumptions are that the population is
sufficiently large and the system connectivity is high {\it i.e.} the
agents display high mobility or they experiment interaction at a 
distance (for instance electronic transactions). 
In this work the population of agents will be fixed to $N_{ag}=1000$
and the number of time steps will be of order $t_f=10^5-10^6$ in such
a way that both assumptions be also consistent with the fact that 
agents have no memory.

Starting from an initial state at $t=0$ taken as $p_k(0)$ chosen at random 
(in the interval [0,1]) for each agent $k$, 
the system evolves by iteration during $t_f$ time steps following these
procedure.

\vspace{2mm}

\begin{itemize}
\item
1) {\it Selection of players:} 
Two agents, located at random positions $i$ 
and $j$, are selected to interact {\it i.e.} to play the PD game. 
\item
2) {\it Playing pairwise PD:}  
The behavior, C or D, of each player $k$
( $k$=$i$ or $k$=$j$ ) is decided generating a random number $r$
and if $p_k(t) > r$ then he plays C and, conversely, if
$p_k(t) < r$ he plays D.
\item
3) {\it Assessment of success:}
Each of the two players compares his utilities $U_k(t)$, which 
is one of the four PD payoffs: $R$, $S$, $T$ or $P$,
with an {\em estimate} $\epsilon_k(t)$ of his expected utilities.
If $U_k(t) \ge \epsilon_k(t)$ ($U_k(t) < \epsilon _k(t)$ ) 
the agent assumes he is doing well (badly)
and therefore its level of cooperation is adequate (inadequate).
\item
4) {\it Probability of cooperation update:}
If player $k$ is doing well he keeps his 
probability of cooperation $p_k(t)$. On the other hand,  
if player $k$ is doing badly he decreases (increases) his 
probability of cooperation $p_k(t)$ if he played C (D) choosing an uniformly 
distributed value between $p_k(t)$ and 1 ( between 0 and $p_k(t)$).

\end{itemize}

\vspace{2mm}

In order to introduce a simple and natural estimate $\epsilon_k(t)$ let us
consider two players $i$ and $j$ which cooperate, at time $t$, with 
probabilities $p_i(t)$ and $p_j(t)$ respectively 
(and defect with probabilities $1-p_i(t)$ and $1-p_j(t)$ ), thus 
the expected utilities for the player $i$, $U^{RSTP}_{ij}(t)$, 
are given by:
\begin{equation}
U^{RSTP}_{ij}(t)  =  R p_i(t) p_j(t) + S p_i(t) (1-p_j(t)) + T (1-p_i(t)) p_j(t) 
+ P (1-p_i(t)) (1-p_j(t)),
\label{eq:Deltacap}
\end{equation}
while the expected utilities for the player $j$, $U^{RSTP}_{ji}$, 
are obtained by interchanging $i$ and $j$ in the above equation.

This implies that, given the average probability of cooperation $\bar{p}(t)$,  
an arbitrary agent, say number $k$, can estimate his average expected utilities 
as:
\begin{equation}
U_k^{RSTP}(\bar{p}(t))  =  R \bar{p}(t)p_k(t) + S p_k(1-\bar{p}(t)) + T (1-
p_k(t)) \bar{p}(t) + P (1-\bar{p}(t))(1-p_k(t)).
\label{eq:Deltacap2}
\end{equation}
However, it turns out that, in general, the value of $\bar{p}$ is unknown by the 
agents. Hence a simpler estimate that can be used agent $k$ for
his expected utilities $\epsilon_k(t)$ is obtained by
replacing in equation (\ref{eq:Deltacap2}) $\bar{p}$(t) by
his own probability of cooperation $p_k(t)$:
$$\epsilon^{RSTP}_k(t) \equiv  R p_k^2(t) + (S+T)p_k(1-p_k(t)) + P(1-p_k(t))^2$$
\begin{equation}
=  (R-S-T+P) p_k^2(t) +(S+T-2P) p_k(t) +P.
\label{eq:LocalEst}
\end{equation}
In other words, agent $k$ adopts the simplest possible extrapolation {\it i.e.} 
that he is a "normal" individual whose probability of cooperation is 
representative of the average value \footnote{Considering more sophisticated 
agents, which have "good information" on the population (for instance the value 
of $\bar{p}$ at time $t$), does not change substantially the main results 
obtained with these naive agents.}. 

The rule each player follows to update his probability of cooperation 
is quite natural and of the type "win-stay" and "lose-shift". That is,
if the player's utilities $U_k$ are larger than his estimate he keeps 
his probability of cooperation. On the other hand, if the utilities are smaller 
than his estimate he changes his probability of cooperation: a) increasing it if 
he played D or b) decreasing it if he played C. From eq. (\ref{eq:LocalEst}) the 
update of $p_k(t) \longrightarrow p_k(t+1)$ is governed by the sign of $U_k(t)-
\epsilon^{RSTP}_k(t)$ {\it i.e.} by the following 
inequations: 
\begin{equation}
(S+T-R-P)p_k^2(t)-(S+T-2P)p_k(t)+\left( \matrix{R \cr S \cr T \cr P}\right) - P 
\;\;\;\; \matrix{>  \cr 
< } \;\; 0 ; 
\label{eq:inequations}
\end{equation}

in the case $>0$ ($<0$) $p_k$ is increased (decreased).

In the next section we will see that the strategy which results from the 
combination of the proposed measure of success and update rule for $p_k$ -the  
steps 3) and 4)- , produces, for a wide variety of payoff matrices, cooperative 
states with $\bar{p}_{eq} > 0$.

Let us end this section with a remark about the problem addressed here and its 
relation with the evolution of cooperation. In this approach, there is no 
competition of different strategies, all the agents follow the same 
universal strategy which does not evolve over time. However, 
the system is adaptive in the sense that the
probabilities of cooperation of the agents do evolve.

\section{RESULTS}

Depending on the payoffs $R, S, T$ and $P$ the system self-organizes, after a 
transient, in equilibrium states with values of 
$\bar{p}_{eq} $ ranging from 0 to 1. 
The equilibrium asymptotic states can be lumped into 
3 groups according to the degre of cooperation attained: {\em highly 
cooperative} ($\bar{p}_{eq}>0.5$), 
{\em moderately cooperative} ($\bar{p}_{eq} \simeq 0.5$) and
of {\em poorly cooperative} ( $\bar{p}_{eq} < 0.5$).
The outcomes for any arbitrary payoff matrix M$^{RSTP}$ can be understood in 
terms of the updating rule for the cooperation probability and the corresponding 
estimate $\epsilon^{RSTP}$ {\it i.e.} from the inequalities 
(\ref{eq:inequations}).

The payoff matrices which imply a dilemma -those which comply with the chain of 
inequalities (\ref{eq:chainPD})- lead either to $\bar{p}_{eq} = \frac{1}{2}$ or 
to $\bar{p}_{eq} = 0$. 
From (\ref{eq:inequations}) it emerges that $\bar{p}_{eq}=0.5$ occurs in the 
case when  $\epsilon^{RSTP}-P$ has no roots in the interval (0,1] ($p=0$ is 
always one of the two roots) and $\bar{p}_{eq}=0$ in the opposite case.

Some other matrices not belonging to class $M_D$ exhibit a tension between C and 
D and give rise to $\bar{p}\simeq \frac{1}{2}$. 
The matrices which do not embody such 
trade-off produce the situations which depart most from 
$\bar{p}_{eq} \simeq \frac{1}{2}$.

It is illustrative to consider, for a moment, the restricted subset of 24 payoff 
matrices obtained from permutation of the four canonical payoff values because 
it covers the three groups with different cooperation levels mentioned above.  
In fact, the system self-organizes into equilibrium states with seven values of 
$\bar{p}_{eq}$: 2 matrices 
(M$^{3501}$ and M$^{3510}$) produce $\bar{p}_{eq}=1$, 2 matrices 
(M$^{1053}$ and M$^{0153}$) produce $\bar{p}_{eq}=0$. The remaining 20 matrices 
produce intermediate values: $\bar{p}_{eq} \simeq 0.72$ (M$^{5301}$) , 
$\bar{p}_{eq} \simeq 0.62$ (M$^{3510}$), $\bar{p}_{eq} \simeq 0.38$ 
(M$^{0135}$), $\bar{p}_{eq} \simeq 0.28$ (M$^{1035}$) and $\bar{p}_{eq}=0.5$ 
(the other 16 matrices and among them the canonical payoff matrix).
The 24 measures are performed over 500 simulations.
Fig. 1 show the average probability of cooperation for different payoff
matrices vs. time for the 50,000 first time steps. 

\begin{center}
\begin{figure}[h]
\centering
\psfig{figure=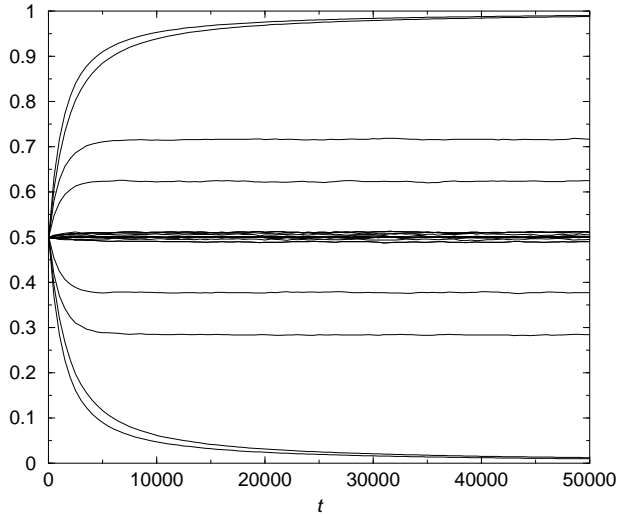,height=7cm}
\caption{Curves of $\bar{p}$ vs. the number of iterations $t$, corresponding to 
the 24 payoff matrices
obtained by permuting the four canonical payoffs $R=3$, $S=0$, $T=5$ 
and $P=1$. The system self-organizes in 
7 different cooperative states with:
$\bar{p}_{eq} =1$,
$\bar{p}_{eq} \simeq 0.72$,
$\bar{p}_{eq} \simeq 0.62$,
$\bar{p}_{eq} \simeq 0.5$, 
$\bar{p}_{eq} \simeq 0.38$, 
$\bar{p}_{eq} \simeq 0.28$ and
$\bar{p}_{eq} \simeq 0$ 
} 
\end{figure}
\end{center}
The mirror symmetry with respect to he value $\bar{p}=0.5$ between the curves 
for $\bar{p}(t)$ corresponding to a given matrix M$^{RSTP}$ and its palindrome 
M$^{PTSR}$ is due to the symmetry of the game when interchanging $R 
\leftrightarrow P$ and $S \leftrightarrow T$ simultaneously with cooperators C 
by defectors D. That is, 

\begin{equation}
\bar{p}^{RSTP}(t)=\overline{(1-p)}^{PTSR}(t).
\label{eq:symmetry}
\end{equation}      

A particular interesting case study is provided by 
payoff matrix M$^{0135}$ with $\bar{p}_{eq} \simeq 0.38$. This result seems, at 
first sight, counter-intuitive: an intermediate cooperation level attained with 
0 reward (and very low sucker's payoff) !
Nevertheless, let us show how the updating rule for the cooperation probability 
explains this outcome. The estimate for this matrix, given by the parabola 
\begin{equation}
\epsilon_k^{0135} = p_k^2 -6p_k +5,
\label{eq:est0135}
\end{equation}
plotted as a solid curve in Fig.2 (the horizontal lines at $S=1$ and $T=3$ cut 
the parabola at abscises $p_S = 3 - \sqrt 5$ and $p_T = 3 - \sqrt 7$, 
respectively). The cooperation update rule tells us the agent $k$ increases his 
probability of cooperation when he plays D and gets $T=3$ if $p_k$ is less than 
$p_T = 3 - \sqrt 7 < 0.5$, {\it i.e.} this temptation is not enough ($T < 
\epsilon_k^{0135}(p_k)$). On the other hand, he decreases his probability of 
cooperation when he plays C and gets $R=0$, independently of the value of $p_k$, 
or when he gets $S=1$ if $p_k$ is less than $p_S = 3 - \sqrt 5 > 0.5$. In the 
remaining situations the player keeps his probability of cooperation. Thus a 
value of $\bar{p}_{eq}$ between 0 and 0.5 is not surprising after all, rather it 
is the result of given the two competing probability of cooperation flows. All 
this analysis for payoff matrix M$^{0135}$ works also for any set of payoffs 
obeying the inequalities:
\begin{equation}
P>T>S>R,
\label{eq:chain0135}
\end{equation}   
the only thing that changes is the value of $\bar{p}_{eq}$. We will come back 
over this particular payoff matrix to illustrate how $p_{eq}$ changes under 
arbitrary variations of the payoffs.

\vspace{4mm}

{\it The Effect of changing payoffs}

\vspace{2mm}

We are now going to analyze the effect of changing the payoff matrix in order to go 
beyond the 24 permutations of the canonical payoffs.

We have seen that the sign of $U_k-\epsilon_k$ controls the update of $p_k$. 
From the definition of $\epsilon_k$, as an estimate of utilities of agent $k$, it is 
clear that it is bounded from above and from below by the largest and smallest of the 
four payoffs, respectively. Thus, $U_k - \epsilon_k$ may have different signs, 
depending on the value of $p_k$, only for the two intermediate payoffs.
Let us denote by $p_1$ the value of $p_k$ such that the estimate
$\epsilon_k$ becomes equal to the larger payoff, $p_2$ the value of $p_k$ such that 
the estimate becomes equal to the second larger payoff, and so on.
Therefore, it is easy to see that the change 
in $\bar{p}_{eq}$ is controlled by the displacements of $p_2$ and $p_3$ (for instance, 
for M$^{0135}$, $p_2 \equiv p_T$ and $p_3 \equiv p_S$). 
If $p_2$ or $p_3$ correspond to the cooperative payoffs, 
$R$ or $S$, then its displacement to the right (left) decreases (increases) 
the proportion of C-agents for whom $U_k>\epsilon_k$ which are, on average, the 
ones who remain C after playing the game. 
This in turn decreases (increases) $\bar{p}_{eq}$. 
On the other hand, if $p_2$ or $p_3$ correspond to the non cooperative 
payoffs, $T$ or $P$, then its displacement to the right (left) decreases 
(increases) the proportion of D-agents for whom $U_k>\epsilon_k$ which are, on 
average, the ones who remain D after playing the game. This in turn increases 
(decreases) $\bar{p}_{eq}$.
  
The payoff matrix M$^{0135}$ will serve to illustrate the effect the changes
in the values of the payoffs have on $p_{eq}$.  
We will proceed by modifying one of the four payoffs at a time and keeping 
fixed the remaining three in such a way that he chain of in-equalities 
(\ref{eq:chain0135}) is preserved.
This variation of a quantity that results when the payoff $X$ is modified and the 
other three payoffs remain fixed is denoted by $\delta_X$. The estimates that result 
from these changes are the curves shown in Fig.2. Let us consider first the changes 
$\delta_{S^+}$, produced by an increment in the sucker's payoff from $S=1$ to $S=2$ 
(which transforms  M$^{0135}$ into M$^{0235}$), $\delta_{S^-}$, produced by a 
decrease from $S=1$ to $S=0$ (which transforms M$^{0135}$ into M$^{0035}$).  
For M$^{0235}$, $p_3$ is the abscise of the point $\epsilon_k \equiv S = 2$ 
(filled up triangle in Fig. 2(A)) and for M$^{0035}$, $p_3$ is the abscise of the 
point $\epsilon_k \equiv S = 0$ (filled down triangle in Fig. 2(A)), while the 
corresponding $p_2$ are the abscises of the points $\epsilon_k \equiv T = 3$ 
(filled triangles: up for M$^{0235}$ and down for M$^{0035}$ in Fig.2(A)).      
We can see that increasing (decreasing) the sucker's payoff, from 
$S=1$ to $S=2$ ($S=0$), produces a displacement of $p_2$ to the right 
(left), from $3-\sqrt 7 \simeq 0.354$ to 0.4 (to $\frac{7-\sqrt 33}{4} \simeq 
0.314$), and of $p_3$ to the left (right), from $3-\sqrt 5 \simeq 0.764$ to 
0.6 (to 1).  
Hence, both changes point in the same direction increasing 
(decreasing) $p_{eq}$ as can be observed in Fig. 3 (dotted lines vs. solid 
lines). In other words, 
$$\delta_{S^+}(p_2-p_3)\simeq (0.4-0.354)-(0.6-0.764)=0.21 >0$$
\begin{equation}
\delta_{S^-}(p_2-p_3)\simeq (0.314-0.354)-(1-0.764)=-0.276 <0
\label{eq:deltaS}
\end{equation}

Similarly, we denote by $\delta_{P^+}$ the variations produced by an increment in 
the punishment, from $P=5$ to $P=6$ (which transforms  M$^{0135}$ into M$^{0136}$), 
and by $\delta_{P^-}$ the variations produced by a decrease in the punishment, 
from $P=5$ to $P=4$ (which transforms M$^{0135}$ into M$^{0134}$).  
For both matrices, the corresponding $p_2$ and $p_3$ are the abscises of  
the points $\epsilon_k \equiv T = 3$ and $\epsilon_k \equiv S = 1$ 
(non filled triangles in Fig.2: up for M$^{0136}$ and down for M$^{0134}$ ), respectively.      
Also in Fig. 2(A) we see that changing the punishment, from $P=5$ to $P=6$ 
($P=4$), produces a displacement of $p_2$ 
to the right (left), from $3-\sqrt 7\simeq 0.354$ to $\frac{4-\sqrt 10}{2} 
\simeq 0.419$ (to 0.25) , and of $p_3$ to the right (left), from 
$3-\sqrt 5 \simeq 0.764$ to  $\frac{4-\sqrt 6}{2} \simeq 0.775$ (to 0.75), 
hence the two changes point in opposite directions: the first tends to increase 
(decrease) $p_{eq}$ and the second to decrease (increase) it. As the first 
displacement is larger it dominates, and the net result is an increase (decrease) of 
$p_{eq}$ as can be observed in Fig. 3 (dot-dashed lines vs. solid line). That is: 
$$\delta_{P^+}(p_2-p_3)\simeq (0.419-0.354)-(0.775-0.764)=0.054>0,$$

\begin{equation}
\delta_{P^-}(p_2-p_3)\simeq (0.25-0.354)-(0.75-0.764)=-0.09 < 0.
\label{eq:deltaP}
\end{equation}

\begin{center}
\begin{figure}[h]\centering
\psfig{figure=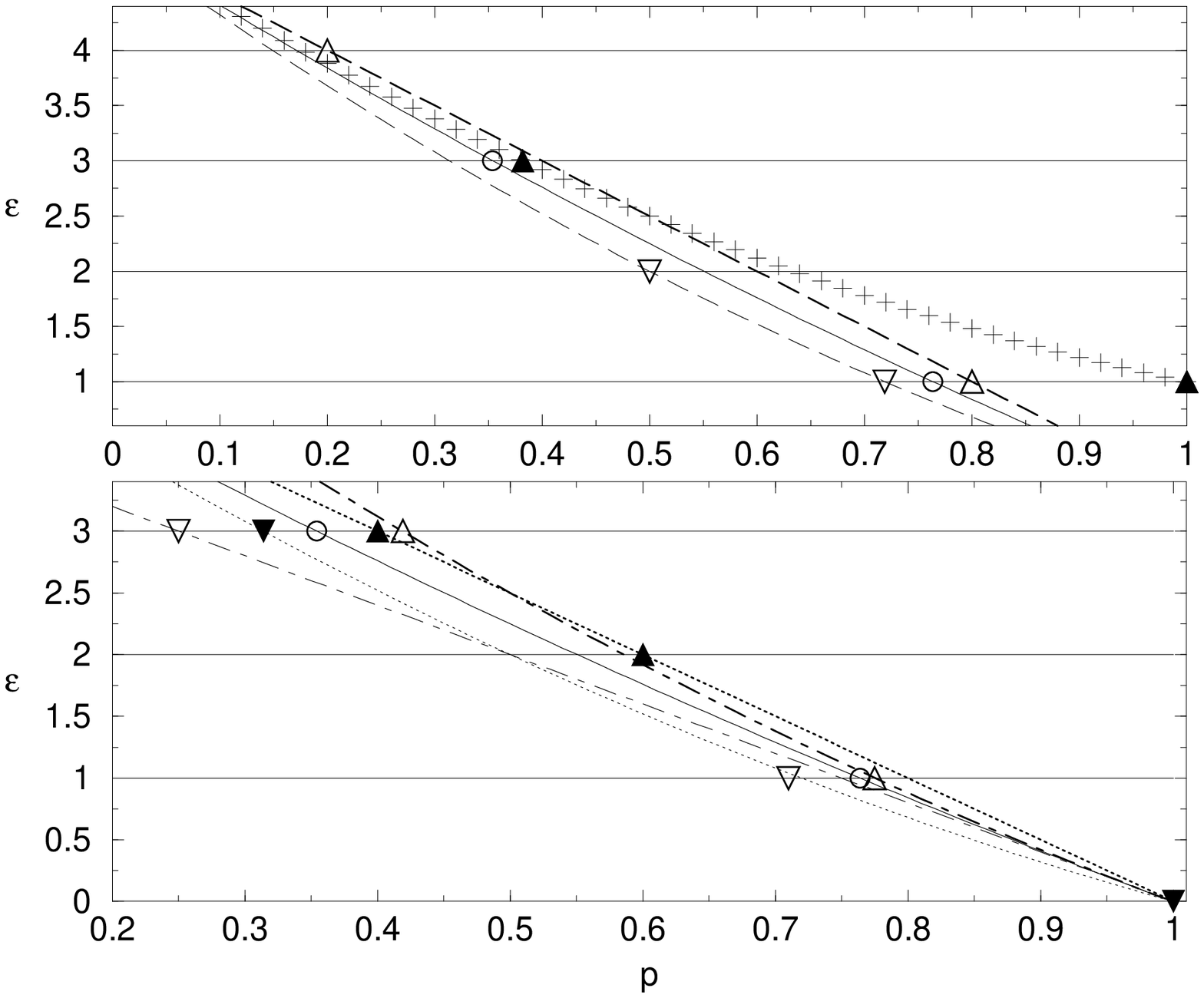,height=9cm}
\caption{(A) Below: The estimate $\epsilon^{0135}(p)$ vs. $p$ (solid line) 
compared with the estimates that result for independent variations of payoffs 
$S$ and $P$, once at a time: $\epsilon^{0035}$ (dotted thin line), $\epsilon^{0235}$ 
(dotted thick line), $\epsilon^{0134}$ (dot-dashed thin line) and $\epsilon^{0136}$ 
(dot-dashed thick line). The circles correspond to the points $\epsilon_{0135}=1$
and $\epsilon_{0135}=3$. The filled up (down) triangles correspond to the points 
$\epsilon_{0235}=2$ and $\epsilon_{0235}=3$ ($\epsilon_{0035}=0$
and $\epsilon_{0035}=3$). The non-filled up (down) triangles correspond to the points 
$\epsilon_{0136}=1$ and $\epsilon_{0136}=3$ ($\epsilon_{0134}=1$ and $\epsilon_{0134}=3$). 
 (B) Above: The estimate $\epsilon^{0135}(p)$ vs. $p$ (solid line) 
compared with the estimates that result for independent variations of payoffs 
$T$ and $R$, once at a time: $\epsilon^{0125}$ (dashed thin line), $\epsilon^{0145}$ 
(dashed thick line) and $\epsilon^{1135}$ (+'s). The circles correspond to the points 
$\epsilon_{0135}=1$ and $\epsilon^{0135}=3$. The filled up triangles correspond to 
the points $\epsilon^{1135}=1$ and $\epsilon^{1135}=3$. The non-filled up (down) 
triangles correspond to the points $\epsilon^{0145}=1$ and $\epsilon^{0145}=4$ 
($\epsilon^{0125}=1$ and $\epsilon^{0125}=2$). See text.} 
\end{figure}
\end{center}

On the other hand, let us consider the variations produced by the increment of the 
temptation $\delta_{T^+}$, from $T=3$ to $T=4$ (which transforms  M$^{0135}$ into 
M$^{0145}$), and by its decrease $\delta_{T^-}$, from $T=3$ to $T=2$ (which 
transforms M$^{0135}$ into M$^{0125}$).  
For M$^{0145}$, $p_2$ is the abscise corresponding to the point $\epsilon_k \equiv 
T = 4$ (non-filled up triangle) and for M$^{0125}$, $p_2$ is the abscise of the 
point $\epsilon_k \equiv T = 2$ (non-filled down triangle), while the corresponding 
$p_3$ are the abscises of the points $\epsilon_k \equiv S = 1$ (non-filled triangles: 
up for M$^{0145}$ and down for M$^{0125}$).         
In Fig. 2(B) we can see that increasing (decreasing) the sucker's payoff, from 
$T=3$ to $T=4$ ($T=2$), produces a displacement of $p_2$ to the left 
(right), from $3-\sqrt 7 \simeq 0.354$ to 0.2 (to 0.5) and of $p_3$ to the 
right (left), from $3-\sqrt 5 \simeq 0.764$ to 0.8 (to $\frac{7-\sqrt 17}{2} 
\simeq 0.719$). Hence both changes point in the same direction decreasing 
(increasing) $p_{eq}$ as can be observed in Fig. 3 (dashed lines vs. solid 
line). That is: 
$$\delta_{T^+}(p_2-p_3)\simeq (0.2-0.354)-(0.8-0.764)=-0.19 <0$$
\begin{equation}
\delta_{T^-}(p_2-p_3)\simeq (0.5-0.354)-(0.719-0.764)=0.19 >0
\label{eq:deltaT}
\end{equation}
With a similar argument one realizes that increasing (decreasing) the reward 
$R=0$  $p_{eq}$ decreases (increases).

In summary, for payoff matrices like M$^{0135}$, which obbey the chain of
in-equalities (\ref{eq:chain0135}), we found two 
expected results: a higher value of $p_{eq}$ can be reached by increasing the   
sucker's payoff $S$ (which makes C-agents more altruistic) or decreasing the temptation 
$T$ (reducing the incentives to free ride). Additionally we found two, {\it a priori}, 
unexpected results: a higher value of $p_{eq}$ can also be reached by increasing the 
punishment $P$ or decreasing the reward $R$. 
By an inspection of Fig. 2(A) the effect of an increment of $P$ can be understood as 
rising the expectations of the D-agents which in turn diminishes the fraction of agents 
that are satisfied after playing the game.
Similarly, from Fig. 2(B) we can see that an decrease of $R$ makes the C-agents 
less ambitious and increase the fraction of altruists.

\begin{center}
\begin{figure}[h]
\centering
\psfig{figure=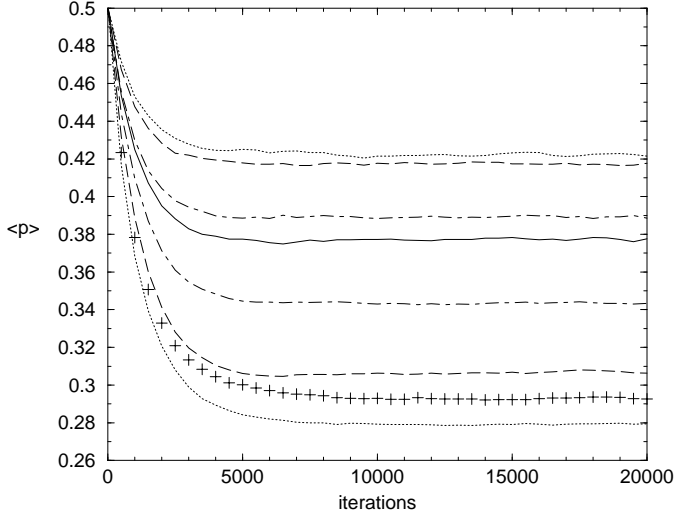,height=7cm}
\caption{The effect of independent variations of payoffs $S,T$ and $P$ (the 
reward remains fixed at $R=0$), once at a time, from payoff matrix M$^{0135}$. 
Variations of $S$: $S=2$ (dotted thick line), $S=0$ (dotted thin line). 
Variations of $P$: $P=6$ (dot-dashed thick line), $P=4$ (dot-dashed thin line).
Variations of $T$: $T=4$ (dashed thick line), $T=2$ (dashed thin 
line). Variations of $R$: $R=1$ ('+').} 
\end{figure}
\end{center}

It is worth remarking that, for the case of payoffs obeying 
(\ref{eq:chain0135}), something which at first seems as innocent as to 
interchange the two non cooperative payoffs $T$ and $P$ has a dramatic 
consequence: it transforms a system with an intermediate level of cooperation 
into one with null cooperation. This can be understood by comparing the estimate 
(\ref{eq:est0135}) for payoff matrix M$^{0135}$ to the one for M$^{0153}$ which 
is given by
\begin{equation}
\epsilon_k^{0153} = -3 p_k^2 +3.
\label{eq:est0153}
\end{equation}
Both estimates have maximum value of $P$ (5 and 3, respectively, at $p_k=0$) but 
the important difference is that in the first case $P$ is the maximum payoff 
while in second one $P<T$. 
Thus in this second case, only the agents which play C can do badly, and then 
the only possible change for $p_k$ (according to its update rule) is a reduction 
till it reaches zero value.    

Finally, let us include a note regarding the efficiency to attain cooperative 
regimes. The state of maximum cooperation $\bar{p}_{eq}=1$ 
is reached for payoff matrices such that $S \geq R > \max \{ T, P \}$ plus the 
condition that equation
\begin{equation}
(S+T-R-P)p^2-(S+T-2P)p+R-P = 0, 
\label{eq:peq1}
\end{equation}
has no roots in the interval [0,1] different from $p=1$ (which is always a root 
of (\ref{eq:peq1})). This condition on the roots is because in the opposite 
case, when there is a root $p_x$ in-between 0 and 1, it follows easily from the 
inequations (\ref{eq:inequations}) that $\bar{p}$ converges to the semi-sum of 
$p_x$ and 1.  
It can be checked by elemental algebra that this is the case of, for instance, 
payoff matrices M$^{3501}$, M$^{3510}$.

\section{CONCLUSIONS}

The success of the strategy to attain cooperative regimes for a wide variety of 
games (payoff matrices) - mainly those which implies dilemmas or clearly favor 
D - relies on the combination of the proposed measure of success and update 
rule for the probability of cooperation. Basically it works by tuning the agent's 
cooperation guided by a trade-off between efficiency (increase of utilities) 
and equity (indirect reciprocity). 
If the agent is doing well he maintains his probability of cooperation otherwise 
he changes it. When he is doing badly playing D he becomes more cooperative, 
{\it i.e.} he increases his probability 
of cooperation attempting to change to behavior C and explore this alternative 
behavior. Conversely, if he is doing badly playing C then he decreases his 
probability of cooperation attempting to change to behavior D and see what 
happen.

An interesting extension of the model would be to allow competition
of different strategies to promote their evolution {\it i.e.} players 
which imitate the best-performing ones in such a way that lower 
scoring strategies decrease in number and the higher scoring increase.

Another possibility would be to allow the use of distinct payoff matrices.
For instance, individuals inclined to cooperate (defect) might 
be represented by agents using the payoff matrix M$^{5301}$ (M$^{1035}$)
while "neutral" ordinary agents by those using the canonical payoff 
matrix M$^{3051}$. 
This would make possible to study if mutants inclined to D can invade a group of 
neutral individuals or individuals inclined to C and drive out all cooperation.

Here I considered a MF approximation which   
neglects all the spatial correlations. One virtue of this 
simplification is that it shows the model does not require
that agents interact only with those within some geographical proximity 
in order to sustain cooperation. Playing with fixed neighbors
is sometimes considered as an important ingredient to
successfully maintain the cooperative regime \cite{cra2001},\cite{nm92}.
However, the quality of this MF approximation
depends on the nature of the system one desires to model, and varies whether one 
deals 
with human societies, viruses \cite{tc99}, cultures of bacteria \cite{d89} 
or market of providers of different products. 
In order to consider situations in which the effect of geographic closeness 
cannot be neglected, an alternative version of this model, might include a 
spatial PD game, 
in which individuals interact only (or mainly) with those within some 
geographical proximity. In that case, the study of spatial patterns seems an 
interesting issue to address. Work is in progress in that direction.

\vspace{2mm}

{\bf Acknowledgments.}

I thank R. Axelrod for valuable opinions and R. Donangelo for useful discussions 
and criticism.


\vspace{2mm}

\begin{thebibliography}{99}


\bibitem{axel84} R. Axelrod, in {\it The Evolution of Cooperation}, 
Basic Books, New York, 1984;
R. Axelrod, in {\it The Complexity of Cooperation}, 
Princeton University Press 1997.
These two volumes include lots of useful references.
Also it is illuminating the Chapter 3 of
{\it Harnesing Complexity} by R. Axelrod and M. Cohen, The Free Press 1999. 

\bibitem{h90} H.A. Simon, Science {\bf 260}, 1665 (1990). 

\bibitem{axel81} R. Axelrod, Am. Polit. Sci. Rev. {\bf 75}, 306 (1981).

\bibitem{g88} J.M. Grieco, Jour. of Politics {\bf 50}, 600 (1988).

\bibitem{ff02} E. Fehr and U. Fischbacher, Econ. Jour. {\bf 112}, 478 
(2002).

\bibitem{cs01} K. Clark and M. Sefton, Econ. Jour., {\bf 111}, 51 (2001).

\bibitem{bf99} I. Bohnet and B. Frey,
Jour. of Economic Behavior and Organization {\bf 38}, 43 (1999).

\bibitem{am93} J. Andreoni and J.H. Miller, Econ. Jour. {\bf 103}, 570 (1993).

\bibitem{bs92} K. Binmore and L. Samuelson, Jour. of Econ. Theo.
 {\bf 57}, 278 (1992).

\bibitem{k89} B. Kogut, Jour. of Industrial Economics {\bf 38}, 183 (1989).

\bibitem{w89} D. Warsh, Harvard Business Review {\bf 67}, 26 (1989).

\bibitem{h01} R. Hausmann, Foreign Policy {\bf 122}, 44 (2001)

\bibitem{g91} J. Goldstein, International Studies Quarterly {\bf 35}, 
195 (1991).

\bibitem{p91} R. Powell, Am. Polit. Sci. Rev. {\bf 85}, 1303 (1991).

\bibitem{s71} G. H. Snyder, International Studies Quarterly
{\bf 15}, 66 (1971).

\bibitem{wn99} L.M. Wahl and M.A. Nowak, Jour. of Theor. Biol. {\bf 200}, 
307 (1999); Jour. Theor. Biol. {\bf 200}, 323 (1999).

\bibitem{ns94} M.A. Nowak and K. Sigmund, Jour. Theor. Biol.{\bf 168}, 
219 (1994).

\bibitem{n90} M.A. Nowak, Theor. Pop. Biol. {\bf 38}, 93 (1990)

\bibitem{md97} M. Mesterton-Gibbons and L.A. Dugatkin, Animal Behaviour
{\bf 54}, 551 (1997).

\bibitem{dmh92} L.A. Dugatkin, M. Mesterton-Gibbons and A.I. Houston. 
Trends in Evolutionary Ecology {\bf 7}, 202 (1992).

\bibitem{n51} J. Nash, Annals of Mathematics {\bf 54}, 286 (1951).

\bibitem{ep98} J. Epstein, 
{\it Zones of Cooperation in Demographic Prisoner's Dilemma} 
Complexity, Vol. 4, Number 2, November-December 1998.  

\bibitem{rg66} A. Rapoport and M. Guyer, {\it General Systems} {\bf 11}, 205
(1966).

\bibitem{df02} R. Donangelo and H. Fort, to be published elsewhere.

\bibitem{cra2001} M.D. Cohen, R.L. Riolo and R. Axelrod, 
Rationality and Society {\bf 13}, 5 (2001).

\bibitem{nm92} M.A. Nowak and R. May, Nature {\bf 359}, 826 (1992)


\bibitem{tc99} P.E. Turner and L. Cho, Nature {\bf 398}, 441 (1999).

\bibitem{d89} R. Dawkins,  {\it The Selfish Gene:  New Edition} 
Oxford University Press (1989).


\end{thebibliography}
\end{document}